# Modeling of Frequency Security Constraints and Quantification of Frequency Control Reserve Requirements for Unit Commitment

Likai Liu, *Student Member, IEEE*, Zechun Hu, *Senior Member, IEEE*

*Abstract*—The high penetration of converter-based renewable energy sources has brought challenges to the power system frequency control. It is essential to consider the frequency security constraints and frequency control reserve requirements in unit commitment (UC). Considering that the risk of frequency insecurity varies under the changeable operational condition, we propose to optimize the PFC droop gains and reserve capacities in the UC model to provide diverse control efforts in different risk levels adaptively. Copula theory is used to establish the joint distribution model among frequency control performance, secondary frequency control (SFC) reserve capacities, and power fluctuations. Then the distributionally robust optimization technique is utilized in the SFC reserve requirement determination to handle the possible error in the probability model. The UC simulation is conducted on IEEE 118-bus system to test the proposed optimal PFC droop gain strategy and SFC reserve requirement quantification method. Simulation results show that the proposed optimal PFC droop gain strategy is better than the traditional fixed PFC droop gain setting on economic efficiency and operational flexibility. Besides, the SFC reserve requirement calculated by the proposed method is more appropriate than the actual SFC reserve capacity in the historical operation.

*Index Terms*--Unit Commitment, frequency security constraints, frequency control reserve requirement, primary frequency control.

## Nomenclature

### Abbreviation

| | |
|---|---|
| FNC | Frequency nadir constraint. |
| DSFR | Demand-side flexible resource. |
| MTDP | Maximum toleratable disturbance power. |
| PFC | Primary frequency control. |
| RoCoF | Rate of change of frequency. |
| SFC | Secondary frequency control. |
| TG | Traditional generator. |
| UC | Unit commitment. |

### Indices & sets

| | |
|---|---|
| $n$ | Index of buses. |
| $t$ | Index of dispatch periods. |
| $\mathcal{I}, \mathcal{J}$ | Set of TGs and converter-based DSFRs. |
| $\mathcal{W}, \mathcal{S}$ | Set of wind power plants and solar power plants. |
| $\mathcal{L}$ | Set of power transmission lines. |

### Parameters

| | |
|---|---|
| $B$ | Susceptance of the transmission line. |
| $C^{SU}$ | Start-up cost of generator. |
| $C_i^{Spin}, C_i^{SFC}$ | Spinning and SFC reserve costs. |
| $D, H$ | Normalized damping and inertia factors of the centralized frequency response model. |
| $F^{Re}$ | High-pressure turbine fraction of the reheat steam turbine. |
| $\Delta f_{max}^{Nadir/QSS}$ | Nadir/quasi-steady-state frequency deviation limit. |
| $K^g, K^b$ | Droop gains of TG and DSFR. |
| $L$ | Power system load power. |
| $\overline{M}_i, \underline{M}_i$ | Upper and lower limits of the PFC droop gain of frequency regulation resource $i$. |
| $P^{max}, P^{min}$ | Maximum and minimum power outputs of TG. |
| $P^{Fore}$ | Forecasted power output of a renewable generation resource. |
| $P^b$ | Base operational point of DSFR. |
| $\Delta P, \Delta f$ | Power imbalance and frequency deviation. |
| $r_{Req}^{Op+/-}$ | Upward/downward operation reserve requirement. |
| $r_{Step}^{SFC}$ | Step size of capacity when calculating SFC reserve requirement. |
| $r_{min}^{SFC}$ | Minimum SFC reserve capacity in historical data. |
| $R^g, R^b$ | Droop factors of TG and DSFR. |
| $RoCoF_{max}$ | Rate of change of frequency maximum limit. |
| $S^{g/b}, S_{base}$ | Capacity of TG/DSFR and the power base value. |
| $T^{Go}, T^{Tu}$ | Governor and turbine time constant of TG. |
| $T^{ON}, T^{OFF}$ | Minimum online and offline times of TG. |
| $T^{Re}$ | Reheater time constant of TG. |
| $T^{Co}$ | Converter time constant of DSFR. |
| $V$ | Ramp rate of a generation resource. |
| $W$ | Capacity of the transmission line. |
| $\phi_{i,k}^g, \varphi_{i,k}^g$ | First-degree/constant term coefficient of the piecewise-linear generation cost function of generator $i$ on the $k^{th}$ segment. |

### Decision variables

| | |
|---|---|
| $P^g$ | Base operational point of TG. |
| $r^{Op+/-}$ | Upward/downward operation reserve capacity. |
| $r^{Spin+/-}$ | Upward/downward spinning reserve capacity. |
| $r^{PFC+}, r^{PFC-}$ | Upward and downward PFC reserve capacities. |
| $r^{SFC+}, r^{SFC-}$ | Upward and downward SFC reserve capacities. |
| $x^g$ | Binary variable representing the ON/OFF status of TG. |
| $x^{PFC}$ | Binary variable indicating whether a regulation resource participates in PFC. |
| $y, z$ | Binary variables representing the start-up and shutdown processes of TG. |
| $\theta$ | Voltage angle of the bus. |

This work was supported by Key Research and Development Program of Inner Mongolia, China Grant 2021ZD0039.(*Corresponding author: Zechun Hu.*)

The authors are with the Department of Electrical Engineering, Tsinghua University, Beijing, 100084, China (e-mail: zechhu@tsinghua.edu.cn).

## I. INTRODUCTION

CARBON neutral has become the consensus of most countries in the world. To achieve this ambitious goal, 33% and 25% of the total power generation should be provided by wind and solar energies by 2050 [1]. The high penetration of renewable generations will lower the system inertia level and intensify the power fluctuation, jeopardizing the frequency quality and increasing the risk of frequency instability [2], [3]. Thus, it is essential to consider the frequency control in power system scheduling.

Unit commitment determines the frequency response model and the most significant power disturbance under *N*-1 contingency [4]. They both have significant influences on the power system frequency security. Thus, considering frequency security constraints in UC has been an active research direction recently [5].

The frequency security constraints require the RoCoF, quasi-steady-state frequency deviation, and frequency nadir within their limits [6], where FNC is highly nonlinear. [7]. To preserve the linearity of the UC problem, several researchers have obtained fruitful results in the linearization of the FNC, which can be divided into model-driven methods [8]–[13] and data-driven methods [14], [15].

Another critical issue is the variable risk of frequency insecurity. The large-scale integration of renewable generations increases the variation ranges of several key parameters in the frequency response model [16], [17], such as the inertial level, resulting in the varying risk of frequency insecurity. The droop gain decides the power regulation from PFC, which has prominent influences on the frequency security. Nevertheless, the droop gains of the PFC resources are currently fixed, which cannot accommodate the changeable risk of frequency insecurity. Thus, a novel strategy is proposed in this paper to optimize the PFC droop gains in the UC model.

Apart from ensuring frequency security by PFC, power systems also need SFC to guarantee the frequency quality. SFC requires a certain reserve to counteract the power fluctuation, while excessive reserve will increase the operational cost. Therefore, properly setting the SFC reserve capacities is also essential to operate the power system safely and economically [18]–[21].

The traditional SFC reserve requirement calculation methods are usually based on operational experience [22]–[25]. The significant improvements of power system informatization and fast developments in data science enable us to research and propose more scientific approaches to determine the SFC reserve requirement.

The historical operation data contain the relationship among power fluctuation, SFC reserve capacity, and frequency control performance, which can be utilized to determine the SFC reserve requirement. Yang *et al.* proposed a method to evaluate the SFC reserve capacity adequacy by calculating the conditional probability of reaching the control performance standard under the available SFC reserve capacity [21]. The fluctuation intensities of load power and renewable generations are time-varying, making the SFC reserve requirement also time-independent. Nevertheless, this method does not utilize information about future power fluctuations. Zhang *et al.* built a multiple linear regression model, which reflects the relationship among the power fluctuations, SFC reserve capacity, and frequency control performance, to calculate the SFC reserve requirement [20]. However, the regression-based method cannot reflect and manage the risk of frequency control standard violation, which is less flexible than the probability-based method. Moreover, the existing methods may face challenges when the historical data are insufficient or inaccurate, because the established relationship may be unfaithful. This situation often happens when the newly-built wind power or solar power plants integrate to grid.

Most of the existing literature only considers the post-fault frequency security constraints or mainly focuses on the frequency control reserve requirement in the UC problem. Nevertheless, both of these two factors will affect UC optimization. Therefore, they should be considered simultaneously in the UC model. Zhang *et al.* made a preliminary study on this problem [26], where the primary frequency response is simplified as a constant ramp function of time, and the SFC reserve capacity is determined without properly considering the frequency control performance.

To compensate for the insufficiency of the existing researches, this research proposes a UC model considering the frequency security constraints and frequency control capacity requirements. The major contributions of this work are summarized as follows:
1) The droop gains of the frequency regulation resources are optimized in the UC model to adaptively provide different PFC services in different risks of frequency insecurity.
2) A novel SFC reserve requirements calculation method is proposed based on the Copula theory and distributionally robust optimization technique.

The remainder of the paper is organized as follows. Section II introduces the basic ideas of this paper. The frequency security constraints are developed in Section III. Section IV gives the data-driven SFC reserve requirement calculation method. The proposed UC model is given in Section V. Section VI presents the numerical experiments. Section VII concludes this paper.

## II. RESEARCH MOTIVATIONS

The power balance in current power systems is typically achieved by three coordinated processes: the day-ahead unit commitment, intra-day economic dispatch, and real-time frequency control. The power fluctuations in traditional power systems are mild because they mainly come from the load demands, leading to a relatively small difficulty in system frequency control. As a result, the power system UC usually takes less consideration of the subsequent real-time frequency control. The high penetration of renewable generations has increased the difficulty of power system frequency control. Thus, the frequency control requirements should be more carefully considered during the power system scheduling process.

## A. Necessity of Considering Frequency Security Constraints in Unit Commitment

UC determines the on-off statuses of TGs, and thus directly decides the system inertia and available PFC resources. Moreover, the most significant power disturbance in power system operation is the dropping of the generator with the largest power output, which also depends on the UC results. These three factors significantly impact the system frequency deviation after the equipment failure, so the operational plan obtained from UC has prominent influences on the post-fault system frequency security. Therefore, it is essential to consider frequency security constraints in UC optimization.

## B. Merit of Optimizing PFC Droop Gains in UC

In addition to the large-scale integration of centralized renewable power generations, the increments of electric vehicles and distributed generations on the demand side are also very fast. The increasing penetrations of these converter-interfaced resources have led to significant heterogeneities in power system generation and load resources. Moreover, their power injections/consumptions have strong randomness and volatility, making the frequency response model highly changeable, significantly influencing the frequency security constraints. For example, a higher risk of the FNC violation exists when the system inertia level is lower than its average value.

The droop gain decides the power regulation value from PFC, which significantly influences the quasi-steady-state frequency and frequency nadir. Nevertheless, the static and fixed droop gains may face challenges to deal with the dynamic-changing frequency response model. This study proposes to combat the dynamic-changing control system with the dynamic-changing PFC droop gain. To be specific, the droop gains of the PFC resources are optimized in the UC model.

The benefits of this strategy lie in two aspects: first, the variable droop gain allows a PFC resource to make different control efforts in different operational risks, which helps to improve the frequency security; second, the variable droop gain can decrease the number of generators which are forced being started to guarantee the frequency security constraints, and thus save the power system operational cost. The PFC droop gains are coupled with the ON/OFF statuses of the generators, so the droop gains should be optimized in the UC stage.

## C. Necessity of Considering Frequency Control Capacity Requirements in Unit Commitment

Power systems need certain frequency control reserve capacities to counteract the power fluctuations. Online TGs provide the majority part of frequency control reserve. The allocation of frequency control reserve capacities and the dispatch of power generation resources are two closely coupled problems, which should be jointly optimized in the UC stage to improve the economic efficiency of power system operation.

The calculations of frequency control reserve requirements are essential to operate the power system safely and economically. Based on the droop gain of a PFC resource and the maximum allowable frequency deviation of the system, the maximum power regulation that needs to be provided by this resource during PFC can be calculated, and PFC reserve requirement can be determined. Nevertheless, SFC reserve requirement determination currently relies on operational experience, which may be unsafe and uneconomical under the largely distinct operating conditions caused by the high proportion of renewable generations. Therefore, the determination of the SFC reserve requirement is also an important research issue.

## III. FREQUENCY SECURITY CONSTRAINTS AND PRIMARY FREQUENCY CONTROL

### A. Frequency Security Constraints

Fig. 1 illustrates a frequency control system comprised of TGs and converter-interfaced DSFRs, such as battery energy storage and electric vehicles. It is assumed that the renewable generation resource neither provides inertia response nor participates in frequency control in this paper. For a TG without a reheater, $F^{Re}$ and $T^{Re}$ are set as 0.

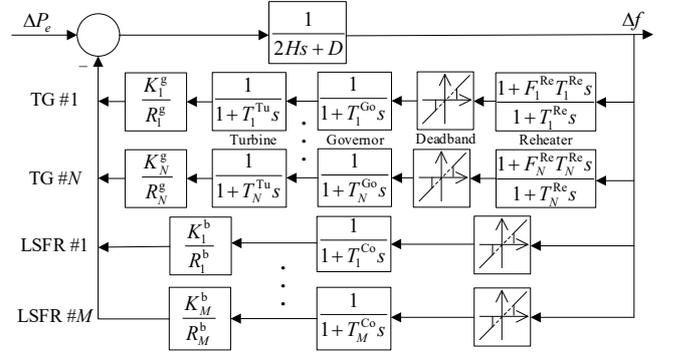

Fig. 1 Centralized frequency control model.

To guarantee frequency security, the power system operator requires some post-fault dynamic frequency metrics must stay within their limitations. These frequency security metrics are introduced as follows.

*1) RoCoF*

In the first few seconds following the large power disturbance, the frequency droop is only arrested by the inertia response from the surviving synchronous units. To avoid triggering the RoCoF relay, the RoCoF in a measurement window should be limited. The RoCoF constraint can be guaranteed by restricting the highest value of RoCoF, which happens at the very instant of the outage.

$$\left| \dot{f}_{\max} \right| = \frac{|\Delta P|}{2H \cdot S_{base}} \leq RoCoF_{\max}, \quad (1)$$

*2) Quasi-Steady-State Frequency Deviation*

A few seconds after the power disturbance, PFC will arrest the frequency decay and then recover it to the quasi-steady-state. The quasi-steady-state frequency deviation and its security constraint are expressed as follows.

$$\Delta f^{QSS} = \frac{|\Delta P|}{S_{base} \cdot (D + R_\Sigma^g + R_\Sigma^b)} \leq \Delta f_{\max}^{QSS}, \quad (2)$$

where $R_\Sigma^g$ and $R_\Sigma^b$ are the integrated droop factors of TGs and DSFRs, as

$$R_\Sigma^g = \frac{1}{S_{base}} \cdot \sum_{i \in \mathcal{I}} \left( S_i^g \cdot \frac{K_i^g}{R_i^g} \right), \quad (3)$$

$$R_\Sigma^b = \frac{1}{S_{base}} \cdot \sum_{j \in \mathcal{J}} \left( S_j^b \cdot \frac{K_j^b}{R_j^b} \right). \quad (4)$$

*3) Frequency Nadir*

The frequency control model shown in Fig. 1 is a high order model, making the frequency nadir formulation very complex. To simplify the model, authors in [13], [27] omit the regulator deadband, the governor and turbine blocks of TG, and assume that $T^{Re} = T \gg T^{Co} \approx 0$. Then, the FNC after a stepwise disturbance $\Delta P_e(s) = -\Delta P/s$ is formulated as (5). The detailed derivations can be found in [13], [27].

$$\left|\Delta f^{nadir}\right| = \frac{|\Delta P|}{S_{base} \cdot (D + R_\Sigma^g + R_\Sigma^b)} \left(1 + \sqrt{\frac{T(R_\Sigma^g - F)}{2H}} e^{-\zeta \omega_n t_m}\right) \leq \Delta f_{max}^{Nadir}, \quad (5)$$

The above FNC is highly nonlinear because the parameters $R_\Sigma^g$, $R_\Sigma^b$, $F$, $H$, $\zeta$, $\omega_n$, and $t_m$ depend on the ON/OFF statuses of generators and the droop gains of the PFC resources, making the power system scheduling model including the FNC becomes a nonlinear programming problem that is hard to solve.

As shown in (5), the frequency nadir is proportional to the power disturbance, implicating that the frequency deviation limitation $\Delta f_{max}^{Nadir}$ corresponds to a $\overline{\Delta P}$. Furthermore, the FNC can be ensured by restricting the TGs' power outputs below the MTDP. The same authors proposed a meticulously designed extreme learning machine-based network to forecast MTDP [15]. In this paper, that model is modified by replacing the binary variable indicating whether a resource participates PFC with its droop gain to adapt for the changeable PFC droop gain strategy.

Based on the MTDP forecast model, the FNC can be formulated as a piecewise linear affine function

$$\min_{1 \leq \ell \leq \mathcal{L}} \left( c_\ell \cdot [\boldsymbol{\Psi}', \boldsymbol{\Phi}', H]^\top + h_\ell \right),$$

where $\ell$ and $\mathcal{L}$ are the index and number of segments of the piecewise function. In the $\ell^{th}$ segment, $c_l$ is a vector composed of the first-degree item coefficients, and $h_l$ is the constant item coefficient. $\boldsymbol{\Psi}'$ and $\boldsymbol{\Phi}'$ are vectors composed of the original and newly-generated characteristic parameters of frequency regulation resources, respectively, and they are linear functions of droop gains of frequency regulation resources.

*B. Primary Frequency Control Constraints*

This study proposes to combat the dynamic-changing control system with the changeable PFC droop gain. Before optimizing the droop gains of PFC resources in the UC optimization model, the following constraints should be added to the UC model:

$$x_i^{PFC} \cdot \underline{M}_i \leq K_i^{g/b} \leq x_i^{PFC} \cdot \overline{M}_i, \quad \forall i \in \{\mathcal{I}, \mathcal{J}\}, \quad (6)$$

$$x_i^{PFC} \leq x_i^{UC}, \quad \forall i \in \mathcal{I}, \quad (7)$$

$$c_\ell \cdot [\boldsymbol{\Psi}'_{\setminus i}, \boldsymbol{\Phi}'_{\setminus i}, H_{\setminus i}]^\dagger + h_\ell \geq \frac{P_{i,t}}{S_{base}}, \quad \forall i \in \mathcal{I}, \ell, t \quad (8)$$

$$\frac{P_{i,t}}{H_t \cdot S_{base} - H_{i,t} \cdot x_{i,t}^{UC} \cdot P_i^{max}} \leq RoCoF^{max}, \quad \forall i \in \mathcal{I}, t \quad (9)$$

$$\frac{P_{i,t}}{\sum_{j \in \{\mathcal{I},\mathcal{J}\}/i} K_j^{g/b} \cdot P_j^{max}/R_j + L_t \cdot D} \leq \Delta f_{max}^{QSS}, \quad \forall i \in \mathcal{I}, t \quad (10)$$

$$r_{i,t}^{PFC+/-} \geq \frac{K_i^{g/b}}{R_i} \cdot P_i^{max} \cdot \Delta f_{max}^{Nadir}, \quad \forall i \in \{\mathcal{I}, \mathcal{J}\}, \forall t \quad (11)$$

Equation (6) restricts the droop gain ranges of PFC resources. Equation (7) restricts that a TG in OFF state cannot participate in the PFC. FNC is guaranteed by equation (8), which requires that after tripping any single generator, the MTDP of the power system composed of other online generators and DSFRs should be larger than the power output of the broken generator. Equation (9) is the RoCoF constraint, (10) is the steady frequency deviation constraint, (11) guarantees the PFC reserve capacity is sufficient even when the frequency deviation reaches its upper limit.

## IV. SFC RESERVE REQUIREMENT DETERMINATION

As introduced before, properly setting the SFC reserve capacity is also essential to operate the power system safely and economically. This section will introduce the proposed SFC reserve requirement calculation method, whose objective is to choose a minimum SFC reserve capacity enough for the AGC system to satisfy the power balancing control performance standard.

*A. Power Balancing Control Performance Standard*

The North American Electric Reliability Corporation has published several standards to measure frequency control performance [28]. In this paper, $A_1/A_2$ standard is used as the control performance standard because the historical data are taken from an actual control area adopting $A_1/A_2$ standard. Whereas, the proposed method also applies to control areas adopting the CPS standard. The definition of $A_1$ is the times of the area control error (ACE) crossing zero within a stipulated timespan, and $A_2$ is the average ACE over this period, as

$$A_2 = \frac{1}{\Gamma} \sum_{\tau=1}^{\Gamma} ACE_\tau, \quad (12)$$

where $\Gamma$ is the timespan, $ACE_\tau$ is the $\tau^{th}$ ACE in this period. $A_1$ only qualitatively reflects whether the ACE crosses zero, while $A_2$ quantitively shows the control performance. Therefore, the quantification of the SFC reserve requirement is based on $A_2$ in this research.

*B. Criteria for Adequacy of SFC Reserve Capacity*

In power systems, the power variations from load power and renewable generations mainly cause frequency fluctuations. In this research, the changes of load power and renewable generations from the start to the end of a time interval are used to describe their variations. The power variation intensities are time-varying. Therefore, to calculate the SFC reserve requirement of a time interval, its power variations should be forecasted first. In this research, the extreme learning machine-based interval prediction method [29] is utilized to forecast the ranges of the power variations. The forecasted power fluctuation intervals are denoted as $[\underline{d}, \overline{d}]$, where $\underline{d}$ is composed of the lower limits of load, wind, and solar power variations, $\overline{d}$ is composed of the upper limits of load, wind, and solar power variations.

Given the forecasted power fluctuation intervals $[\underline{d}, \overline{d}]$, the SFC reserve capacity $r$ should be selected to guarantee that the conditional probability of compliance of frequency control standard, under the predicted power fluctuations and chosen SFC reserve capacity, is larger than the preset confidence coefficient $\alpha$, as

$$\mathbb{P}(|A_2| \leq A_2^* | \underline{d} \leq d \leq \overline{d}, r - \Delta r \leq r^{\text{SFC}} \leq r + \Delta r) \geq \alpha \quad (13)$$

where $A_2^*$ is the threshold value specified in the control performance standard, $d$ is the set composed of the load, wind, and solar power variations, $\Delta r$ is used to form the range of SFC reserve capacity, $r^{\text{SFC}}$ is the historical SFC reserve capacity.

*C. Copula-Based Joint Distribution Model*

The high-dimensional probability in (13) cannot be accurately calculated through making statistics on historical data. To solve this problem, the conditional probability model (13) is built by utilizing the Copula theory. Because errors are inevitable when constructing the probability model, distributionally robust optimization techniques are used to calculate the SFC reserve requirement based on the built probability model.

Considering that the upward SFC reserve requirement is not related to the downward SFC reserve capacity and vice versa, the joint distribution of $A_2$, $d$, $r^{\text{SFC+}}$ and that of $A_2$, $d$, $r^{\text{SFC-}}$ are built separately. To avoid the repetitive expression, we use $r^{\text{SFC}}$ to denote $r^{\text{SFC+}}$ or $r^{\text{SFC-}}$ in the following descriptions.

Copula theory is a kind of effective tool to build multivariate distribution [30]. It transforms the construction of joint cumulative distribution into modeling the marginal distributions and fitting the Copula function. By utilizing the Copula theory, the joint cumulative distribution function (CDF) of $A_2$, $d$, and $r^{\text{Sec}}$ is shown as

$$\mathcal{F}_{A_2,d,r^{\text{SFC}}}(A_2,d,r^{\text{SFC}}) = \mathcal{C}_{A_2,d,r^{\text{SFC}}}(\mathcal{F}_{A_2}(A_2),\ldots,\mathcal{F}_{d_S}(d_S),\ldots,\mathcal{F}_{r^{\text{SFC}}}(r^{\text{SFC}})) \quad (14)$$

where $\mathcal{F}$ is the cumulative distribution function, $\mathcal{C}$ is the Copula function, $\mathcal{F}_{A_2}(A_2)$, $\mathcal{F}_{d_S}(d_S)$, and $\mathcal{F}_{r^{\text{SFC}}}(r^{\text{SFC}})$ are the marginal CDFs of $A_2$, $d_S$, and $r^{\text{SFC}}$, respectively. The joint probability density function (PDF) $f_{A_2,d,r^{\text{SFC}}}$ is the derivative of the joint CDF $\mathcal{F}_{A_2,d,r^{\text{SFC}}}$ on $A_2$, $d$, and $r^{\text{SFC}}$, as

$$f_{A_2,d,r^{\text{SFC}}}(A_2,d,r^{\text{SFC}}) = \frac{\partial \mathcal{F}_{A_2,d,r^{\text{SFC}}}(A_2,d,r^{\text{SFC}})}{\partial A_2 \partial d \partial r^{\text{SFC}}}$$
$$= \varsigma_{A_2,d,r^{\text{SFC}}}(\mathcal{F}_{A_2}(A_2),\ldots,\mathcal{F}_{d_S}(d_S),\ldots,\mathcal{F}_{r^{\text{Sec}}}(r^{\text{SFC}})) \quad (15)$$
$$\cdot f_{A_2}(A_2) \cdot \prod_{d \in d} f_d(d) \cdot f_{r^{\text{SFC}}}(r^{\text{SFC}})$$

The conditional PDF of $A_2$ under $d$ and $r^{\text{SFC}}$ can be calculated by dividing the joint PDF $f_{d,r^{\text{SFC}}}(d,r^{\text{SFC}})$ by the joint PDF $f_{A_2,d,r^{\text{SFC}}}(A_2,d,r^{\text{SFC}})$, as

$$f_{A_2|d,r^{\text{SFC}}}(A_2|d,r^{\text{SFC}}) = \frac{f_{A_2,d,r^{\text{SFC}}}(A_2,d,r^{\text{SFC}})}{f_{d,r^{\text{Sec}}}(d,r^{\text{SFC}})}$$
$$= \frac{\varsigma_{A_2,d,r^{\text{SFC}}}(\mathcal{F}_{A_2}(A_2),\ldots,\mathcal{F}_{d_S}(d_S),\ldots,\mathcal{F}_{r^{\text{Sec}}}(r^{\text{SFC}})) \cdot f_{A_2}(A_2)}{\varsigma_{d,r^{\text{SFC}}}(\mathcal{F}_{d_L}(d_L),\ldots,\mathcal{F}_{r^{\text{Sec}}}(r^{\text{SFC}}))} \quad (16)$$

The joint PDF $f_{d,r^{\text{SFC}}}(d,r^{\text{SFC}})$ can be built in a similar way as (14)-(15).

There are many different functions that can be used as the Copula function $\mathcal{C}$ to build the joint distribution. This study chooses the most commonly used functions, including Gaussian Copula, Student-t Copula, Clayton Copula, Gumbel Copula, and Frank Copula, to build the joint distribution (14). After that, the Bayesian information criterion (*BIC*) [31] is utilized to assess the fitting performances of different models, as

$$BIC = -2\ln(\hbar) + q * \ln(\mathcal{X}), \quad (17)$$

where $q$ is the number of parameters that need to be fitted in the Copula function, $\hbar$ represents the value of the maximum likelihood function [32], and $\mathcal{X}$ indicates the number of data points used in building the model.

*D. Distributionally Robust Chance Constraint*

The historical data may be insufficient or inaccurate when newly integrated renewable generation resource exists, which will make the probability model estimated based on the historical data unfaithful. Furthermore, this may lead to the calculated SFC reserve requirement being incorrect. To handle the possible errors in the probability model, chance constraint (13) is reformulated as a distributionally robust chance constraint (DRCC), which requires that the chance constraint is satisfied for all the PDFs within the ambiguity set constructed from the samples [33], as

$$\min_{\mathbb{Q} \in \widehat{\mathcal{P}}} \mathbb{P}^{\mathbb{Q}}(|A_2| \leq A_2^* | \underline{d} \leq d \leq \overline{d}, r - \Delta r \leq r^{\text{SFC}} \leq r + \Delta r) \geq \alpha \quad (18)$$

In this research, the ambiguity set $\widehat{\mathcal{P}}$ is built based on the Wasserstein metric. The distance between distributions $\mathbb{Q}_1$ and $\mathbb{Q}_2$ defined by the Wasserstein metric is:

$$\mathcal{D}_W(\mathbb{Q}_1,\mathbb{Q}_2) = \sup_{f \in \mathbb{L}} \left\{ \int_\Xi f(\xi) \mathbb{Q}_1(d\xi) - \int_\Xi f(\xi) \mathbb{Q}_2(d\xi) \right\}, \quad (19)$$

where $\xi$ is the stochastic variable belonging to the value space $\Xi \triangleq \{\xi \in \mathbb{R} : \mathcal{B}\xi \leq \mathcal{H}\}$, $\mathbb{L}$ denotes the spaces of all Lipschitz functions with $|f(\xi) - f(\xi')| \leq |\xi - \xi'|$ for all $\xi_1, \xi_2 \in \Xi$. The ambiguity set can be expressed as

$$\widehat{\mathcal{P}} \triangleq \{\mathbb{Q} : \mathcal{D}_W(\widehat{\mathbb{P}},\mathbb{Q}) \leq \varepsilon\}, \quad (20)$$

where $\widehat{\mathbb{P}}$ is the empirical distribution constructed from the data set composed of historical data and sample data taken from (16) conditioned on the forecasted power variation intervals and the SFC reserve capacity range $[r - \Delta r, r + \Delta r]$. $\varepsilon$ is the radius of the ambiguity set.

Conditional value at risk (CVaR) is a risk measure that quantifies the expected loss over the part of distribution beyond the confidence level [34]. The DRCC (18) is transformed by utilizing the CVaR approximation [35], as

$$\max_{\mathbb{Q} \in \widehat{\mathcal{P}}} \mathbb{E}^{\mathbb{Q}} \left[ \max \left( \frac{A_2 - A_2^* - \delta^0}{1-\alpha}, \frac{-A_2 - A_2^* - \delta^0}{1-\alpha}, 0 \right) + \delta^0 \right] \leq 0. \quad (21)$$

Then, (21) can be reformulated to a set of constraints according to Corollary 5.1 in [33], as

$$\begin{cases} \lambda\varepsilon + \dfrac{1}{\mathcal{U}}\sum_{\mu=1}^{\mathcal{U}}\sigma_\mu \le 0, \\ \dfrac{A_{2,\mu}-A_2^*+\alpha\delta^0}{1-\alpha}+\gamma_\mu^1\cdot(\mathcal{H}-\mathcal{B}\cdot A_{2,\mu}) \le \sigma_\mu, & \forall \mu, \\ \dfrac{-A_{2,\mu}-A_2^*+\alpha\delta^0}{1-\alpha}+\gamma_\mu^2\cdot(\mathcal{H}-\mathcal{B}\cdot A_{2,\mu}) \le \sigma_\mu, & \forall \mu, \\ \delta^0+\gamma_\mu^3\cdot(\mathcal{H}-\mathcal{B}\cdot A_{2,\mu}) \le \sigma_\mu, & \forall \mu, \\ \left\|\mathcal{A}^\top\gamma_\mu^1-\dfrac{1}{1-\alpha},\ \mathcal{A}^\top\gamma_\mu^2+\dfrac{1}{1-\alpha},\ \mathcal{A}^\top\gamma_\mu^3\right\|_\infty \le \lambda, \forall \mu, \\ \gamma_\mu^1,\gamma_\mu^2,\gamma_\mu^3 \ge 0, & \forall\mu, \end{cases} \quad (22)$$

where $\lambda$, $\sigma_\mu$, $\gamma_\mu^1$, $\gamma_\mu^2$, $\gamma_\mu^3$ are auxiliary variables, $\mu$ and $\mathcal{U}$ are the index and the total number of the data samples.

The set of constraints (22) can be utilized to check whether a SFC reserve capacity $r$ is sufficient to fulfill the frequency control standard. The flowchart of the SFC reserve requirement calculation is shown in Fig. 2. The upward and downward SFC reserve capacity requirements should be calculated separately.

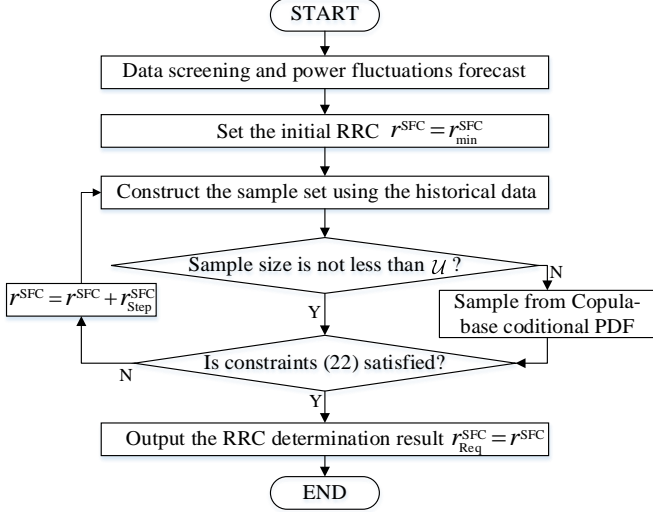

Fig. 2 The flowchart of SFC reserve capacity calculation.

## V. Modified Unit Commitment Model

The UC optimization model considering the frequency security constraints and frequency control capacity requirements is developed in this section.

### A. Objective Function

The objective function of the UC model is to minimize the total operational cost, which is composed of the generation and start-up costs of TGs, as well as reserve costs of all frequency regulation resources.

$$\text{Min.}\ \sum_{t\in\mathcal{T}}\left\{\sum_{i\in\mathcal{I}}\left(C_i^{\text{SU}}y_{i,t}+C_{i,t}^{\text{Gen}}\right)+\sum_{i\in\{\mathcal{I},\mathcal{J}\}}C_i^{\text{Spin}}\left(r_{i,t}^{\text{Spin+}}+r_{i,t}^{\text{Spin-}}\right)\right\} \quad (23)$$

The generation cost of TG $i$ can be formulated as follows:

$$C_{i,t}^{\text{Gen}} = \max_{k\le K}\left(\phi_{i,k}^{\text{g}}\cdot P_{i,t}^{\text{g}}+\varphi_{i,k}^{\text{g}}\right), \quad \forall i\in\mathcal{I}, t. \quad (24)$$

### B. Constraints

The operational constraints in the UC model are presented as follows:

#### 1) Power Flow Constraints

$$\sum_{i\in\mathcal{I}_n}P_{i,t}^{\text{g}}+\sum_{i\in\mathcal{J}_n}P_{i,t}^{\text{b}}-L_{n,t}=\sum_{m:(n,m)\in\mathfrak{I}}B_{n,m}(\theta_{n,t}-\theta_{m,t}), \quad \forall n,t \quad (25)$$

$$B_{n,m}(\theta_{n,t}-\theta_{m,t}) \le W_{n,m}, \quad \forall(n,m)\in\mathfrak{I},t \quad (26)$$

Equation (25) is the DC power flow constraint, where $m:(n,m)\in\mathfrak{I}$ denotes the buses connected with bus $n$, and $\mathfrak{I}$ is the set of all branches. It is assumed that the base operational points of DSFRs are known and cannot be scheduled. Equation (26) is the line capacity constraint.

#### 2) Generator Start-Up and Shutdown Constraints

$$y_{i,t} \ge x_{i,t}^{\text{g}}-x_{i,t-1}^{\text{g}}, \quad \forall i\in\mathcal{I},t \quad (27)$$

$$z_{i,t} \ge x_{i,t-1}^{\text{g}}-x_{i,t}^{\text{g}}, \quad \forall i\in\mathcal{I},t \quad (28)$$

$$\sum_{\tau=t-T_i^{\text{ON}}}^{t-1}x_{i,\tau}^{\text{g}} \ge z_{i,t}T_i^{\text{ON}}, \quad \forall i\in\mathcal{I},t \quad (29)$$

$$\sum_{\tau=t-T_i^{\text{OFF}}}^{t-1}(1-x_{i,\tau}^{\text{g}}) \ge y_{i,t}T_i^{\text{OFF}}, \quad \forall i\in\mathcal{I},t \quad (30)$$

Equations (27) and (28) are the start-up and shutdown logical constraints for TGs. Equations (29) and (30) are the minimum online and offline time constraints of TGs.

#### 3) Operational Range Constraints

$$P_{i,t}^{\text{g/b}}+r_{i,t}^{\text{Spin+}} \le P_i^{\max}x_{i,t}^{\text{g}}, \quad \forall i\in\{\mathcal{I},\mathcal{J}\},t \quad (31)$$

$$P_{i,t}^{\text{g/b}}-r_{i,t}^{\text{Spin-}} \ge P_i^{\min}x_{i,t}^{\text{g}}, \quad \forall i\in\{\mathcal{I},\mathcal{J}\},t \quad (32)$$

Equations (31)-(32) restrict that the base operational points of the TGs and DSFRs plus/minus the upward/downward spinning reserve capacities are within their power ranges.

#### 4) Ramp Rate Constraints

$$P_{i,t}-P_{i,t-1} \le x_{i,t}V_i+y_{i,t}P_i^{\min}, \quad \forall i\in\mathcal{I},t \quad (33)$$

$$P_{i,t-1}-P_{i,t} \le x_{i,t-1}V_i+z_{i,t}P_i^{\min}, \quad \forall i\in\mathcal{I},t \quad (34)$$

Equations (33) and (34) are the ramp rate constraints between the adjacent base dispatch points of the TG, where the start-up and shutdown are considered. The ramp rate constraints for DSFR are not considered because their ramp rates are fast.

#### 5) Reserve Capacity Constraints

$$r_{i,t}^{\text{Spin+/-}} \ge r_{i,t}^{\text{PFC+/-}}+r_{i,t}^{\text{SFC+/-}}+r_{i,t}^{\text{Op+/-}}, \quad \forall i\in\{\mathcal{I},\mathcal{J}\},t \quad (35)$$

$$\sum_{i\in\{\mathcal{I},\mathcal{J}\}}r_{i,t}^{\text{SFC+/-}} \ge r_{\text{Req},t}^{\text{SFC+/-}}, \quad \forall t \quad (36)$$

$$\sum_{i\in\{\mathcal{I},\mathcal{J}\}}r_{i,t}^{\text{Op+/-}} \ge r_{\text{Req},t}^{\text{Op+/-}}, \quad \forall t \quad (37)$$

For each resource, the spinning reserve capacity should be larger than the sum of PFC, SFC, and operation reserve capacities, as (35). The sum of upward/downward SFC reserve capacity provided by all resources should be larger than the upward/downward SFC reserve capacity requirement calculated by utilizing the method proposed in Section IV. The system operation reserve requirement also need be satisfied, as (37). The system operation reserve requirement is taken as a certain percentage of the system total load power.

#### 6) Primary Frequency Control Constraints

The PFC constraints (6)-(11) should also be included in the UC optimization model.

*Remarks 1*: The forecast errors of the renewable generations are not considered in this UC model because it is not the focus of this study. Nevertheless, there are tremendous methods to handle this issue, and they can be easily introduced into this UC model [36]–[39].

TABLE I GENERATOR PARAMETERS

| Generator (# bus) | 12 | 26 | 89 | 59,61 | 65,66 | 10,69, 80 | 25,49, 100 | 31,46, 87 | 54,103, 111 |
|---|---|---|---|---|---|---|---|---|---|
| Reserve cost ($/MWh) | 7.2 | 9 | 9.6 | 7.8 | 9.6 | 9 | 7.8 | 6 | 7.2 |
| Start up cost ($) | 120 | 400 | 800 | 240 | 500 | 600 | 300 | 80 | 100 |
| Ramp rate (MW/min) | 2 | 7 | 13 | 4 | 8 | 10 | 5 | 1.5 | 1.8 |
| Min. ON/OFF time (h) | 2 | 6 | 8 | 4 | 8 | 6 | 4 | 2 | 2 |

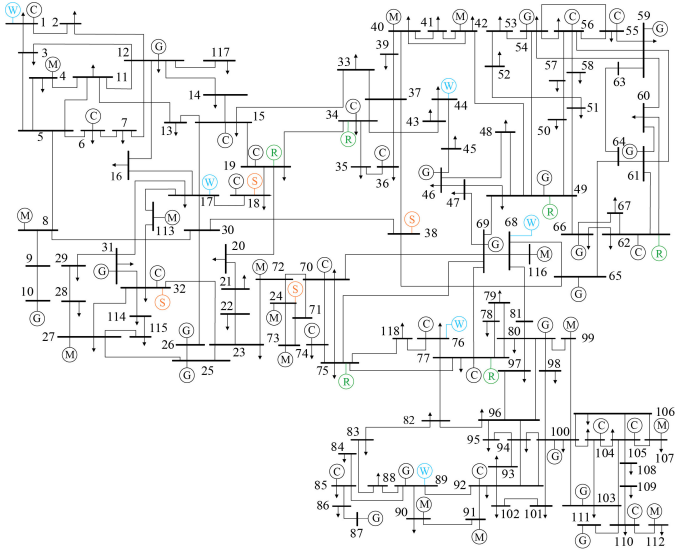

Fig. 3 Diagram of the modified IEEE 118-bus system.

## VI. CASE STUDY

### A. Case Settings

The IEEE 118-bus system is used as the test system, as shown in Fig. 2. We modified it by connecting six wind farms to buses 1, 44, 68, 17, 76, 89 with installed capacities of 400 MW, 300 MW, 300 MW, 330 MW, 400 MW, 300 MW, and four photovoltaic plants to buses 18, 24, 32, 38 with installed capabilities of 400 MW, 400 MW, 330 MW, 300 MW. The generator parameters in the frequency response model are taken from [40], and other generator parameters are listed in Table I. The parameters of DSFRs are shown in Table II. The PFC droop gain range of TG is from 0.5 to 1, and the droop gain range of DSFR is from 0 to 2.

TABLE II DSFR CONVERTER PARAMETERS

| DSFR (# bus) | 19 | 34 | 49 | 62 | 75 | 77 |
|---|---|---|---|---|---|---|
| Capacity (MW) | 40 | 30 | 30 | 33 | 40 | 30 |
| Time constant (s) | 0.01 | 0.01 | 0.01 | 0.01 | 0.01 | 0.01 |
| Reserve cost ($/MW) | 4 | 4 | 4 | 5 | 5 | 5 |
| Droop factor | 0.15 | 0.15 | 0.15 | 0.15 | 0.15 | 0.15 |

The proposed method needs historical data on AGC control performance, SFC reserve capacities, load power, and renewable generations. In this paper, the data from a practical power system in North China are used, and all the historical data are scaled down for the simulations. The historical data from 2016 to 2018 with a total of 821 days are randomly split into training and test sets at a ratio of 6 to 4. The training set is used to train the extreme learning machine-based forecast model and build the Copula-based joint distribution model.

### B. Copula Model Selection

Five Copula functions are applied in Section III.B to construct the joint distribution model of frequency regulation performance, SFC reserve capacity, and power variations. The joint distribution models of each hour are built separately, because the correlation is time-varying. And the joint distribution model will be built for upward and downward SFC reserve separately. The fitting performances of different Copula functions are shown in Table III, where the average fitting performances on 24 hours are calculated.

The joint distribution model established by utilizing Student-t Copula function has the lowest BIC, so we utilize it to build the Copula-based joint distribution model.

TABLE III PERFORMANCES OF DIFFERENT COPULA MODELS

| SFC reserve | Gaussian | Student-t | Gumbel | Clayton | Frank |
|---|---|---|---|---|---|
| Upward | -1139.17 | -1140.15 | -4.58 | -48.88 | -15.69 |
| Downward | -1181.14 | -1185.67 | 7.98 | 7.98 | 7.98 |

### C. Comparison of Different Methods

The proposed UC model is compared with the other two UC models to demonstrate its superiority. The strategies of these three models are shown in Table IV.

TABLE IV STRATEGIES OF DIFFERENT MODELS

| Model | Model A | Model B | Model C |
|---|---|---|---|
| PFC droop gain | Changeable | Fixed | Changeable |
| FNC linearization method | Proposed | Proposed | Reference [13] |
| SFC reserve capacity | Proposed | Proposed | Historical Data |

#### 1) Operational Cost

The comparison results on the operational costs of different models are shown in Fig. 4 and Table V, which gives the ratios of model B and C's operational costs to model A's cost. It can be found from Table V that the generation costs of different models are close, but the reserve cost of model A is the smallest, and its average operational cost is also the smallest. Fig. 4 also shows that the operational cost of model A is smaller than those of the other two models on most test days. These results prove that the proposed UC model is more economically efficient. The average cost is calculated on the test days which are all feasible under three models.

Some test cases are infeasible due to there being no operation plans that satisfy all constraints. The ratio of the infeasible case is utilized to compare the operational flexibilities of different models. The changeable PFC droop gain strategy increases the feasible region of the power system operation, and some infeasible cases due to the violation of frequency security constraints become feasible after adopting the proposed strategy. Therefore, the ratio of the infeasible case of model A is smaller than that of model B. The following analysis will show that both the FNC and SFC reserve requirement of the model A are less conservative than those of model C, which explains why the ratio of the infeasible case of model A is smaller than that of model C.

TABLE V COMPARISON OF THE OPERATIONAL COSTS AND OPERATIONAL FLEXIBILITIES OF DIFFERENT MODELS

| Index | Model A | Model B | Model C |
|---|---|---|---|
| Generation cost ($10^6$ $) | 1.01 | 1.01 | 1.01 |
| Reserve cost ($10^6$ $) | 0.16 | 0.20 | 0.19 |
| Total cost ($10^6$ $) | 1.17 | 1.21 | 1.20 |
| Infeasible case ratio (%) | 1.5% | 4.5% | 10.4% |

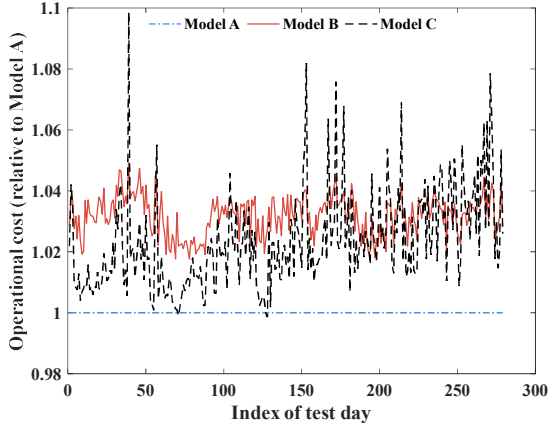

Fig. 4 Operational costs of different methods in the test set.

*2) Primary Frequency Control Strategy and FNC*

The PFC reserve costs of the three models are compared as Fig. 5 and Table VI. The PFC reserve cost of model A is smaller than that of model B, proving that the strategy of optimal PFC droop gain helps save the reserve cost. The frequency bias factors of different methods are also compared. The frequency bias factor is the most important parameter reflecting the PFC characteristic of a system. The frequency bias factor of model A is smaller than that of model B, which proves that the strategy of fixed PFC droop gain is more conservative than the strategy of changeable PFC droop gain.

Both the PFC reserve cost and frequency bias factor of model C are higher than those of model A, which proves that the FNC linearization method proposed by the same authors is less conservative than the method proposed in [13].

TABLE VI  COMPARISONS OF THE PFC RESERVE COSTS AND FREQUENCY BIAS FACTORS

| Model | Model A | Model B | Model C |
|---|---|---|---|
| PFC reserve cost ($10^5$ \$) | 1.00 | 1.35 | 1.08 |
| Frequency bias factor (MW/0.1 Hz) | 67.33 | 74.93 | 75.87 |

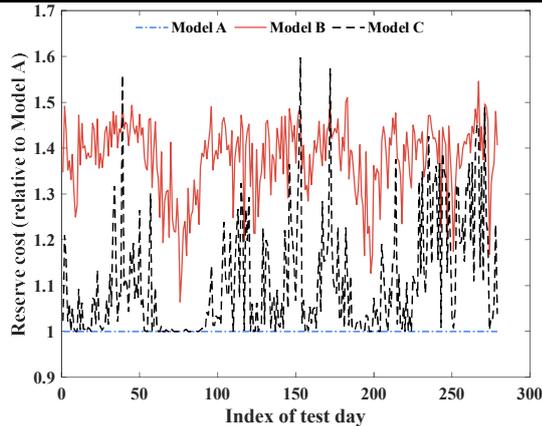

Fig. 5 Comparison on reserve costs of three models.

To test the accuracy of the established linear FNC, the operation plans obtained from different models are used as the inputs, and the frequency control system shown in Fig. 1 is utilized to simulate whether the frequency will exceed the limitation after the failure of any single generator. It is found that the FNC is not violated in any operation plans. In other words, both the FNC linearization methods proposed by the same authors in [41] and Zhang et al. in [13] can guarantee the FNC after $N$-1 failure of the generator.

*3) SFC Reserve Requirement*

The proposed SFC reserve requirement calculation method is tested offline. As shown in Table VII, in the whole test set, the average SFC reserve requirements calculated by the proposed method are smaller than the average SFC capacities in the actual historical data.

Furthermore, the historical time intervals whose actual SFC reserve capacities are close to their calculated SFC reserve requirements are selected to reflect the frequency control performance of the proposed method. The probability of $|A_2|$ exceeding the specified value (40 MW) is calculated on the selected data set and all historical data in the test set, as shown in Fig. 6. The probabilities under the proposed method are no larger than the specified limitation (10%) for all hours. In comparison, the probabilities of historical data are higher than the limitation in many hours, which illustrates that the proposed method helps to improve the frequency control performance.

These results show that the SFC reserve requirement calculated by the proposed method is more appropriate than the actual SFC reserve capacity in the historical operation.

TABLE VII  COMPARISON OF THE OPERATIONAL COSTS OF DIFFERENT MODELS

| Method | Upward SFC capacity (MW) | Downward SFC capacity (MW) |
|---|---|---|
| Proposed | 154.45 | 179.55 |
| Historical | 244.64 | 265.54 |

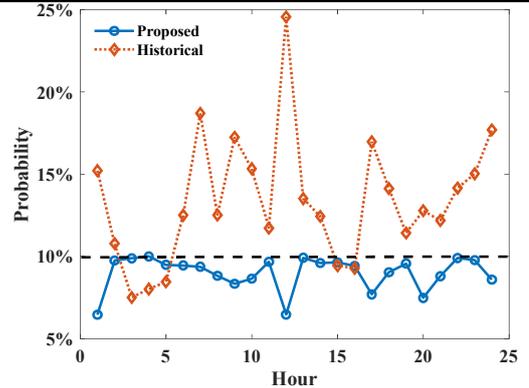

Fig. 6 Comparison on probability of $|A_2|$ exceeding the specified value.

### D. Computation Efficiency

The computation efficiencies of different methods are tested in this subsection. The experiments are conducted on a PC with an Intel i7-7700 CPU 3.6 GHz and 16 GB of memory. The GUROBI solver is utilized to solve the MILP problem [42]. The average computation times of the three models on 10 test days are calculated and shown in Table VIII. It can be found that the computation time of model A is the smallest, and the computation times of all three methods are acceptable in practice.

TABLE VIII  COMPARISON OF THE COMPUTATION TIMES OF THREE MODELS

| Model | Model A | Model B | Model C |
|---|---|---|---|
| Computation time (s) | 40.469 | 112.998 | 122.301 |

## VII. CONCLUSION

This paper proposed a UC model considering the frequency security constraints and frequency control reserve requirements. The frequency security constraints on RoCoF, frequency nadir, and quasi-steady-state frequency deviation are considered. The optimal PFC droop gain strategy is proposed to adopt for the changeable risks of frequency constraint violation. A novel SFC reserve requirements calculation method is proposed by combining the Copula theory and distributionally robust optimization technique. The simulation results conducted on IEEE 118-bus system show that:

1) The optimal PFC droop gain strategy can lower the reserve cost and increase the feasible region of the security-constrained UC problem compared with the fixed PFC droop gain setting adopted in the current operation.
2) The SFC reserve requirement calculated by the proposed method is more appropriate than the actual SFC reserve capacity in the historical operation, which helps to achieve better frequency control performance and save the SFC reserve capacity.

In this research, the modeling of the DSFR is comparatively rough. In future work, the specific models of diverse DSFRs will be utilized in the proposed UC model.